\documentclass[preprint,superscriptaddress,aip,jcp]{revtex4-1}
\usepackage[T1]{fontenc}
\usepackage[latin9]{inputenc}
\usepackage{amsmath}
\usepackage{graphicx}
\usepackage{esint}
\begin{document}

\title{Molecular free energy profiles from force spectroscopy experiments
by inversion of observed committors}

\author{Roberto Covino}
\email{roberto.covino@biophys.mpg.de}
\affiliation{Department of Theoretical Biophysics, Max Planck Institute of Biophysics,
60438 Frankfurt am Main, Germany.}

\author{Michael T. Woodside}
\affiliation{Department of Physics, University of Alberta, Edmonton, Alberta,
T6G 2E1, Canada.}

\author{Gerhard Hummer}
\affiliation{Department of Theoretical Biophysics, Max Planck Institute of Biophysics,
60438 Frankfurt am Main, Germany.}
\affiliation{Institute of Biophysics, Goethe University Frankfurt, 60438 Frankfurt am Main,
Germany.}

\author{Attila Szabo}
\affiliation{Laboratory of Chemical Physics, National Institute of Diabetes and
Digestive and Kidney Diseases, National Institutes of Health, Bethesda,
Maryland 20892-0520, USA.}

\author{Pilar Cossio}
\email{pilar.cossio@biophys.mpg.de}
\affiliation{Department of Theoretical Biophysics, Max Planck Institute of Biophysics,
60438 Frankfurt am Main, Germany.}
\affiliation{Biophysics of Tropical Diseases Max Planck Tandem Group, University
of Antioquia, Medellin, Colombia.}

\begin{abstract}
In single-molecule force spectroscopy experiments, a biomolecule is
attached to a force probe via polymer linkers, and the total extension
\textendash{} of molecule plus apparatus \textendash{} is monitored
as a function of time. In a typical unfolding experiment at constant
force, the total extension jumps between two values that correspond
to the folded and unfolded states of the molecule. For several biomolecular
systems the committor, which is the probability to fold starting from
a given extension, has been used to extract the molecular activation
barrier (a technique known as ``committor inversion''). In this
work, we study the influence of the force probe, which is much larger
than the molecule being measured, on the activation barrier obtained
by committor inversion. We use a two-dimensional framework in which
the diffusion coefficient of the molecule and of the pulling device
can differ. We systematically study the free energy profile along
the total extension obtained from the committor, by numerically solving
the Onsager equation and using Brownian dynamics simulations.
We analyze the dependence of the extracted barrier on the linker stiffness,
molecular barrier height, and diffusion anisotropy, and thus, establish
the range of validity of committor inversion. Along the way, we showcase
the committor of 2-dimensional diffusive models and illustrate how
it is affected by barrier asymmetry and diffusion anisotropy. 
\end{abstract}
\maketitle

\section{Introduction}

In single-molecule pulling experiments, mechanical force is used to
induce conformational transitions in biomolecules \textbf{\cite{Greenleaf2007,Neuman2008}}.
Suppose that the molecule of interest undergoes repeated folding and
unfolding transitions under constant force. The molecular extension,
\emph{i.e.}, the end-to-end distance of the molecule, would then jump
between smaller and larger values. The interpretation of the resulting
time series would be simple if the molecular extension could be directly
monitored experimentally. In this hypothetical case, the folding and
unfolding force-dependent transition rates of the molecule could be
directly obtained by counting the number of transitions per unit time.
In addition, by binning this trajectory one could determine the probability
density of the extension, the logarithm of which is the free energy
profile of the molecule, a procedure known as Boltzmann inversion.
Alternatively, from the trajectory one could determine the probability
that a molecule with a specific extension folds before it unfolds.
This quantity describes the most probable ``fate'' of the system
at any given point of the trajectory, and is known as committor, splitting
probability, or $p_{\mathrm{fold}}$.\cite{Onsager1938} Assuming that the dynamics is
diffusive, the height and shape of the free energy barrier could be
found by differentiation of the committor, a procedure known as committor
inversion.\cite{Chodera2011,Manuel2015}

\begin{figure}
\centering\includegraphics[width=10cm]{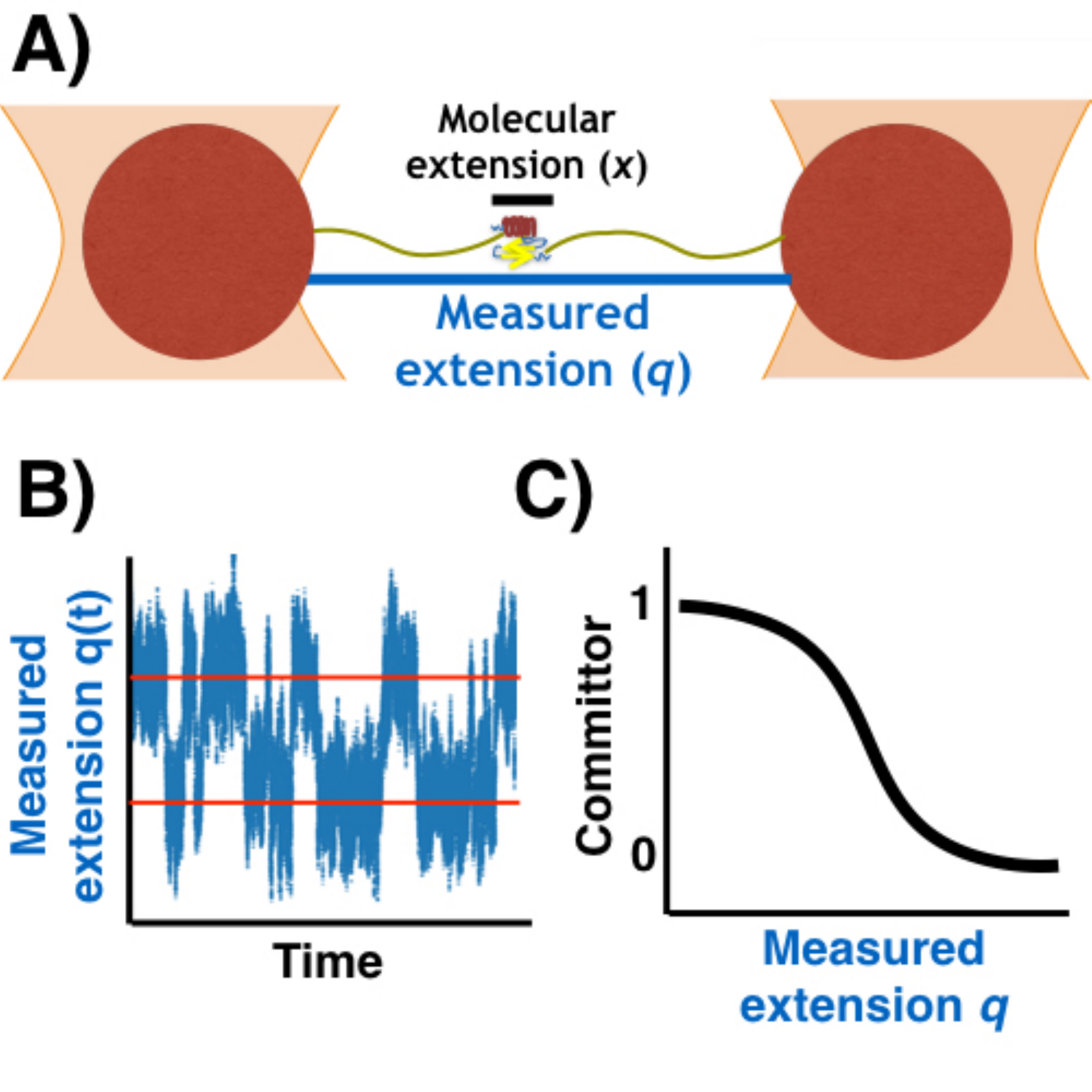}

\caption{\label{fig:1} Schematic of a single-molecule pulling experiment.
\textbf{A}) Example of the experimental setup in force spectroscopy
using optical tweezers. A small biomolecule is attached via polymer
linkers to two beads trapped by laser beams. The molecular extension
$x$ is hidden within the observed extension $q$. \textbf{B) }Trajectory
of the measured extension $q(t)$ as a function of time. \textbf{C)
}The committor estimated from the observed trajectory of the measured
extension $q(t)$ in the interval between the red lines. }
\end{figure}

In reality, however, one cannot directly monitor the molecular extension
itself because the experimental observable is actually the position
of the force probe attached to the molecule by long polymer linkers.
In the case of optical trapping
measurements (Fig. \ref{fig:1}A), for example, the measured extension
($q$) is the extension of the molecule ($x$) plus that of the linkers
attaching the molecule to  mesoscopic beads trapped by laser beams.
What one measures is the time dependence of the inter-bead distance,
yielding a trajectory of the total extension of molecule and linkers
(Fig. \ref{fig:1}B). The free energy profile obtained by Boltzmann
inversion of the observed trajectory of the total extension is a convolution
of the molecule and linker profiles. If the properties of the linker
are known, one can obtain the free energy profile of the molecule
by deconvolution.\cite{Woodside2006} This methodology requires large
amounts of data and works best with low molecular barriers and it
can thus be challenging to use in practice.

As a viable alternative, the group of one of us investigated the free
energy profile obtained by committor inversion of the measured trajectory.\cite{Manuel2015}
If the dynamics of the total extension could be described as diffusion  
on the free energy profile obtained by Boltzmann inversion, then both
the committor and Boltzmann inversion would give the same result.
Consequently, one would still have to use deconvolution to obtain
the molecular profile. However, unless the response of the apparatus
is much faster than that of the molecule, the dynamics of the
total extension cannot necessarily be described as a one-dimensional
diffusive process.\cite{Hummer2010} Committor inversion may therefore
lead to a different free energy profile than Boltzmann inversion.
Experimentally, committor inversion has been applied to DNA hairpin
folding, successfully recovering the free energy profile obtained
from deconvolution of the Boltzmann inverted one.\cite{Manuel2015} It
has also been used to extract free energy barriers encountered when
bacteriorhodopsin is pulled out of a membrane. \cite{Yu2017} This
procedure has the potential to become widely used 
as a viable alternative to 
deconvolution of the profile obtained by Boltzmann inversion.
However, the range of validity of committor inversion in light of
the limitations imposed by probe/linker attachments to the molecule
has not been investigated.

Committor inversion yields the exact molecular free energy profile
in the limit of very stiff polymer linkers (\emph{i.e.}, when the linker
force constant is much larger than those of the molecular extrema).
However, for such linkers the free energy profile obtained from Boltzmann
inversion is already quite close to the molecular one. Moreover, in this
limit the measured
transition or hopping rates become proportional to the diffusion coefficient
not of the molecule, but that of the probe (\emph{e.g.}, mesoscopic beads) attached to
the molecule.\cite{Makarov2014}\textbf{ }Consequently, here we shall
primarily consider soft linkers for which the measured rates are meaningful.
As first pointed out by Thirumalai and coworkers, free energy profiles
are most easily found using stiff linkers, but reliable estimates
of the hopping rates can only be made by using flexible handles.\textbf{\cite{Hyeon2008}}

We will investigate whether transition path theory can aid
the reconstruction of molecular free energy profiles from the information
encoded in the committor estimated from observed trajectories. This will be done in the framework
of a simple model where the molecular and total extensions diffuse
anisotropically on a two-dimensional free energy surface. We previously
used such surfaces to determine the influence of the mesoscopic pulling
device on the observed rates and transition paths.\cite{Cossio2015,Cossio2018}
Here, we obtain the committor both by analyzing Brownian dynamics
trajectories of the total extension \textendash{} as in experiments
\textendash{} and by numerically solving the Onsager equation.\cite{Onsager1938}
Additionally, we derive and validate analytic expressions for the committor obtained
in the high-barrier limit. We then investigate how the extracted barriers
depend on the stiffness of the linker, the shape of the molecular
free energy profile, and the diffusion anisotropy. We find that although
in some realistic cases this procedure yields useful estimates of
the heights of molecular barriers, it is challenging to establish 
its validity in many other cases of practical interest.

\section{Theory }

\begin{figure}
\centering\includegraphics[width=0.6\textwidth]{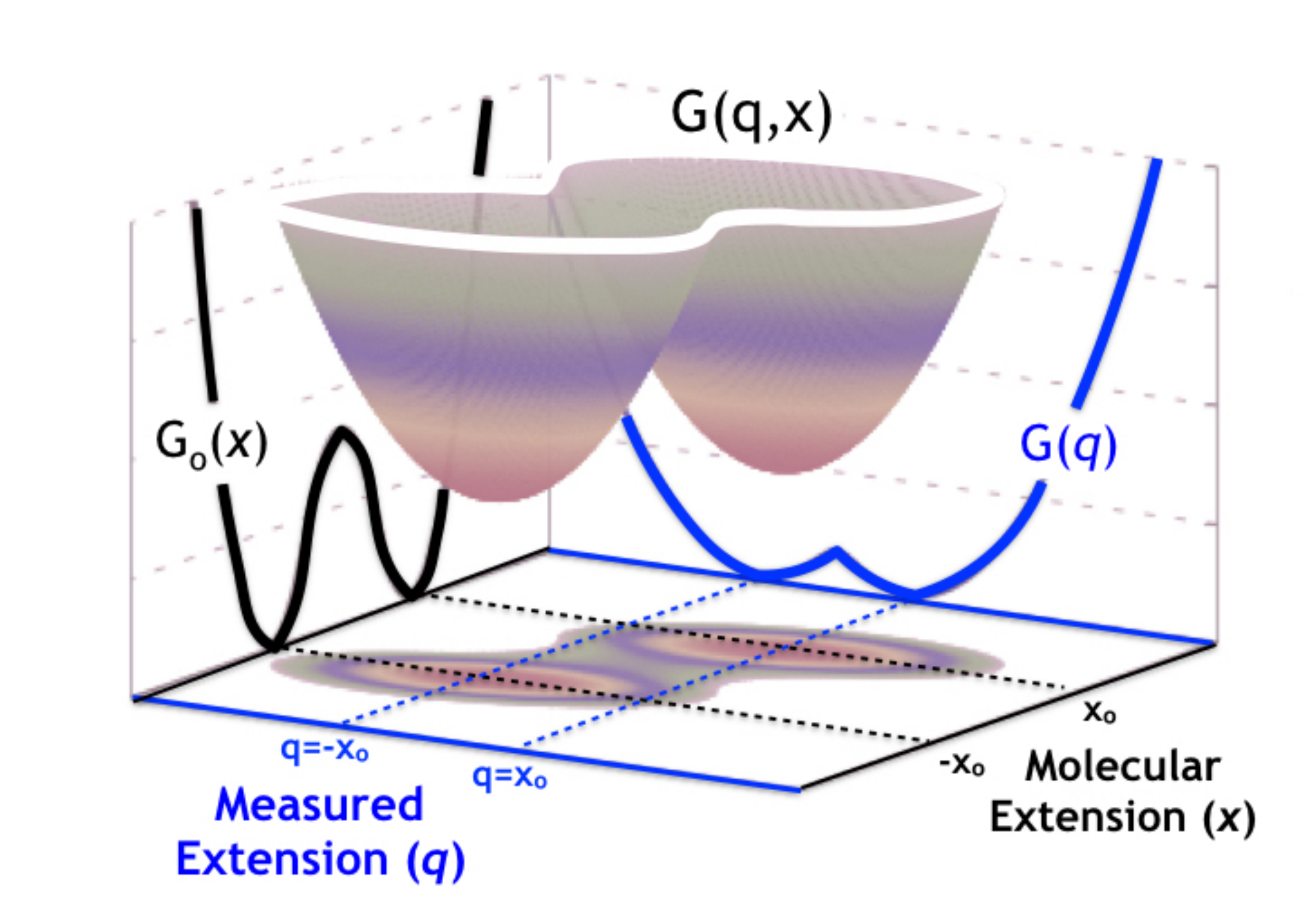}

\caption{\label{fig:2}\textbf{ }Two-dimensional potential surface $G(q,x)$.
We assume that $G_{o}(x)$ (black solid line) is symmetric about its
maximum at $x=0$ and has minima at $-x_{0}$ and $x_{0}$. The potential
of mean force along $q$, $G(q),$ is shown as blue solid line. The
extensions are shifted by constants so that the barrier occurs at
zero.}
\end{figure}

Let $x$ be the molecular (hidden) extension and $q$ be the total
(observable) extension (Fig. \ref{fig:1}A). Let a constant force
be exerted on the system so that the resulting free energy surface
has the form

\begin{equation}
G(q,x)=G_{o}(x)+\frac{\kappa_{l}}{2}(x-q)^{2}.\label{eq:2d energy surface}
\end{equation}
Here, the first term on the r.h.s. is the molecular free energy in
the presence of force, and the second describes the coupling due to
a harmonic linker with spring constant $\kappa_{l}$. For the sake
of simplicity, we will assume the constant force to be subsumed in
$G_{o}(x)$, which is symmetric about its maximum at $x=0$ and has
two minima, corresponding to metastable states, at $x_{1}=-x_{0}$
and $x_{2}=x_{0}$ (Fig. \ref{fig:2}). It is straightforward to generalize
the results presented below to an asymmetric $G_{o}(x)$ and to anharmonic
(\emph{e.g.}, worm-like chain) linkers, albeit at the expense of complicating
the analytical expressions.

We assume that the dynamics on the surface in Eq. \ref{eq:2d energy surface}
is 
diffusive, with position-independent diffusion coefficients $D_x$ and
$D_q$ along the $x$ and $q$ coordinates, respectively. The value of $D_{q}$ is essentially
determined by the Stokes-Einstein diffusion coefficient of the beads
in a laser tweezer experiment and for large beads may thus be slower
than $D_{x}$. By simulating Brownian dynamics one can obtain long
trajectories describing the evolution of the system on the two-dimensional
surface in Eq. \ref{eq:2d energy surface}. The system will spend
most of the time in one of the two metastable states, rarely but rapidly
jumping from one to the other. We can now mimic the typical situation
of force spectroscopy experiments, and assume that only the component
$q(t)$ of the simulated trajectory is observable (see Fig.
\ref{fig:1}B for an example trajectory). From such trajectories one can then
calculate the "observed" committor
$\phi(q)$ (Fig. \ref{fig:1}C), defined as the probability
of reaching the folded minimum before the unfolded minimum, starting
from a given value of $q$.

Alternatively, one can obtain the exact $\phi(q)$ as a conditional
equilibrium average of the two-dimensional committor, $\phi(q,x)$,
which can be accurately obtained by solving the two-dimensional
Onsager equation \cite{Onsager1938} on a grid with the appropriate
boundary conditions (see Methods). Specifically, the observed
committor $\phi(q)$ is given by

\begin{equation}
\phi(q)=\frac{\int_{-\infty}^{\infty}\mathrm{d}x\,\phi(q,x)e^{-\beta G(q,x)}}{\int_{-\infty}^{\infty}\mathrm{d}x\,e^{-\beta G(q,x)}},\label{eq:meancommittor}
\end{equation}
where $\beta=1/k_{\mathrm{B}}T$, $k_{\mathrm{B}}$ is the Boltzmann's
constant, and $T$ the absolute temperature. The denominator in Eq.
\ref{eq:meancommittor} is the exponential of the free energy profile $G(q)$
along $q$, given within a constant by

\begin{equation}
e^{-\beta G(q)}=\int_{-\infty}^{\infty}\mathrm{d}x\,e^{-\beta G(q,x)}=\int_{-\infty}^{\infty}\mathrm{d}x\,e^{-\beta(G_{o}(x)+\kappa_{l}(q-x)^{2}/2)}.\label{eq:pmfq}
\end{equation}
$G(q)$ can be obtained from the observed trajectory by Boltzmann
inversion. If the linker spring-constant is known then $G_{o}(x)$
can in principle be obtained from $G(q)$ by deconvolution,\cite{Woodside2006}
which amounts to a numerically challenging inverse Weierstrass transform.\cite{Hummer2010}

For a one-dimensional diffusive process on $G_{o}(x)$ with
position-independent diffusion coefficient, the committor
$\phi_{o}(x)$ is
given by

\begin{equation}
\phi_{o}(x)=\frac{\int_{x}^{x_{o}}\mathrm{d}y\,e^{\beta G_{o}(y)}}{\int_{-x_{o}}^{x_{o}}\mathrm{d}y\,e^{\beta G_{o}(y)}}.\label{eq:committor1d}
\end{equation}
Thus, by differentiating both sides with respect to $x$, one can
obtain the following inversion formula \cite{Manuel2015}

\begin{equation}
\beta[G_{o}(x)-G_{o}(x_{0})]=\ln\left[\frac{\phi_{o}'(x)}{\phi_{o}'(x_{0})}\right],\label{eq:G_inv-1d}
\end{equation}
which expresses the molecular free energy profile in the interval
$-x_{o}\leq x\leq x_{o}$ in terms of the derivatives of the committor,
denoted as primes. Note that $\phi(-x_{0})=1$ and $\phi(x_{0})=0$
by definition.

For multidimensional diffusive dynamics, there is no analytic relation
between the free energy surface and the committor analogous to Eq.
\ref{eq:G_inv-1d}. Nevertheless, one can formally use this relation
to define a new free energy profile $G_{\mathrm{CI}}(q)$ (CI=committor inversion)
using the committor $\phi(q)$ obtained from the experimental trajectory
in the interval $-x_{o}\leq q\leq x_{o}$ :
\begin{equation}
\beta[G_{\mathrm{CI}}(q)-G_{\mathrm{CI}}(q_{0})]=\ln\left[\frac{\phi'(q)}{\phi'(q_{0})}\right].\label{eq:G_CI}
\end{equation}
Since the dynamics along $q$ cannot be in general described by one-dimensional
diffusion,\cite{Hummer2010,Cossio2015} then in general\cite{Chodera2011}
$G_{\mathrm{CI}}(q)\neq G(q)$. In other words, the committor-inverted and
Boltzmann-inverted profiles are not necessarily the same. It has been
conjectured \cite{Manuel2015} that, in fact, $G_{\mathrm{CI}}(q)$ is very
similar to the hidden molecular free energy profile $G_{o}$ in the
barrier region. This assumption does not have any obvious theoretical
justification, and in the following we will systematically explore
its validity.

We shall now derive approximate analytic expressions for $\phi(q)$
and $G_{\mathrm{CI}}(q)$ when the molecular free energy is symmetric and has
a high barrier. We begin with the calculation of $G(q)$. When the
barrier of $G_{o}(x)$ is high, the major contribution to the integral
in Eq. \ref{eq:pmfq} comes when $x$ is near to the minima of $G_{o}(x)$,
which are located at $x=\pm x_{o}$. Thus, one can approximate the
integral from $-\infty$ to $\infty$ as a sum of two integrals, one
around $x_{1}=-x_{o}$ and the other around $x_{2}=x_{o}$. Then,
we expand $G(q,x)$ in Eq. \ref{eq:meancommittor} around $x_{i}$
($i=1,2$) to second order as $G_{i}(q,x)\approx G(q,x_{i})+(x-x_{i})G'(q,x_{i})+(x-x_{i})^{2}G''(q,x_{i})/2$,
where the primes denote derivatives with respect to $x$. By extending
the range of integration in both integrals to $(-\infty,\infty)$,
and evaluating the resulting Gaussian integrals, we find (to within
a constant)

\begin{equation}
e^{-\beta G(q)}=e^{-\beta\kappa(q+x_{o})^{2}/2}+e^{-\beta\kappa(q-x_{o})^{2}/2},
\end{equation}
where $1/\kappa=1/\kappa_{l}+1/G_{o}''(x_{o})$ and $G_{o}''(x_{o})=G_{o}''(-x_{o})>0$. If we choose the
constant in the definition of $G(q)$ so that $G(q=\pm x_{o})=0$ then

\begin{equation}
\beta G(q)=\ln\left[\frac{\cosh\left(\beta\kappa x_{o}^{2}\right)}{\cosh\left(\beta\kappa x_{o}q\right)}\right]+\frac{\beta}{2}\kappa\left(q^{2}-x_{o}^{2}\right).\label{eq:G_q}
\end{equation}
Let us now evaluate the integral in Eq. \ref{eq:meancommittor} that
determines $\phi(q)$ in an analogous way. We break the integral into
two parts, one around $-x_{o}$ and the other around $x_{o}$, and
expand $G(q,x)$ about these points to second order as before. We
then approximate the committor around $-x_{o}$ as $\phi(q,x)=1$
and set $\phi(q,x)=0$ in the integral around $x_{o}$. Evaluating
the resulting Gaussian integrals, we find that 
\begin{equation}
\phi(q)=\frac{1}{1+e^{-2\beta\kappa x_{o}q}},\label{eq:phiq}
\end{equation}
for $-x_{o}\leq q\leq x_{o}$. This approximate expression is valid
for high molecular barriers and soft linkers. In this regime, $\phi\left(q\right)$
does not depend on the diffusion anisotropy and, more importantly,
it has no direct dependence on the molecular barrier height or shape
(although indirect effects from correlations between the well curvature
and barrier height may occur).

Using Eq. \ref{eq:phiq} and Eq. \ref{eq:G_CI} and requiring that
$G_{\mathrm{CI}}(q=\pm x_{o})=0$ we find that 
\begin{equation}
\beta G_{\mathrm{CI}}(q)=2\ln\left[\frac{\cosh(\beta\kappa x_{o}^{2})}{\cosh(\beta\kappa x_{o}q)}\right],
\end{equation}
which using Eq. \ref{eq:G_q} for $G(q)$ can be rewritten as 
\begin{equation}
G_{\mathrm{CI}}(q)=2G(q)+\kappa\left(x_{o}^{2}-q^{2}\right).\label{eq:G_CI_G_q}
\end{equation}
These approximate expressions are valid for sufficiently large molecular
barriers and sufficiently soft linkers. In this regime of soft linkers
and high molecular barrier, $G_{\mathrm{CI}}(q)$ contains no explicit information
about the molecular barrier height and shape, as determined by $G_{0}(x)$.

\section{Methods}

\subsection{Free energy surfaces}

We model force spectroscopy experiments at constant force as a diffusive
process on the two-dimensional free energy surface $G(q,x)$, given
by Eq. \ref{eq:2d energy surface} (Fig. \ref{fig:2}A), where $q$
and $x$ are the total and molecular extension, respectively, $G_{o}(x)$
is the molecular free energy in the presence of force, and $\kappa_{l}$
is the linker stiffness. We used two analytic forms of the molecular
free energy. A symmetric potential is given by a bistable matched-harmonic
with $G_{o}(x)=\Delta G_{o}^{\ddagger}f\big(x/x^{\ddagger}\big)$,
where 
\begin{equation}
f\big(x\big)=\begin{cases}
-2x^{2} & 0\le|x|\le1/2\\
2(|x|-1)^{2}-1 & 1/2<|x|
\end{cases},\label{eq: parabola matched}
\end{equation}
$\Delta G_{o}^{\ddagger}$ and $x^{\ddagger}$ are the activation
barrier and the distance to the transition state, respectively, in
the presence of force. An asymmetric potential is given by the negative
logarithm of a linear combination of two Gaussian distributions, 
\begin{equation}
\beta G_{o}^{\mathrm{asym}}(x)=-\ln\left(\frac{w}{\sqrt{2\pi s_{1}^{2}}}e^{-(x+x_{0})^{2}/2s_{1}^{2}}+\frac{1-w}{\sqrt{2\pi s_{2}^{2}}}e^{-(x-x_{0})^{2}/2s_{2}^{2}}\right),\label{eq:Gasym}
\end{equation}
where $s_{1}$, $s_{2}$, and $x_{0}$ are the Gaussian widths and
centers, respectively. In particular, we considered a potential displaying
a small barrier by using parameters $s_{1}=0.15$, $s_{2}=1$, and
$w=0.4$; and a potential displaying a larger barrier by using $s_{1}=0.2$,
$s_{2}=0.6$, and $w=0.5$. In both cases the minima are located at
$\pm x_{0}$, with $x_{0}=1.5$.

\subsection{Two-dimensional committor using the Onsager equation}

The Onsager equation \cite{Onsager1938} for a $n$-dimensional diffusive
process $\mathbf{z}$ is 
\begin{equation}
\nabla\cdot\mathbf{D}\left(\mathbf{z}\right)\exp\left[-\beta G\left(\mathbf{z}\right)\right]\nabla\phi\left(\mathbf{z}\right)=0,
\end{equation}
where $\mathbf{D}\left(\mathbf{z}\right)$ is a position-dependent
diffusion tensor. If we assume a two-dimensional diffusion on the
free energy surface $G\left(q,x\right)$, with a diagonal and position-independent
diffusion tensor, then the Onsager equation becomes

\begin{equation}
-D_{x}\partial_{x}\beta G\left(q,x\right)\partial_{x}\phi\left(q,x\right)+D_{x}\partial_{x}^{2}\phi\left(q,x\right)-D_{q}\partial_{q}\beta G\left(q,x\right)\partial_{q}\phi\left(q,x\right)+D_{q}\partial_{q}^{2}\phi\left(q,x\right)=0.\label{eq: 2d adjFP}
\end{equation}
After discretizing this equation, an iterative relaxation method can
provide an accurate numerical solution $\phi\left(q,x\right)$. We
thus consider a mesh on the plane $\left(q,x\right)$ such that both
continuous variables take $N+1$ and $M+1$ discrete values respectively,
$q_{i}\equiv i\Delta q$ and $x_{j}\equiv j\Delta x$, with $\Delta q=\left(q_{\mathrm{max}}-q_{\mathrm{min}}\right)/N$
and $\Delta x=\left(x_{\mathrm{max}}-x_{\mathrm{min}}\right)/M$.
We evaluate the committor on the mesh, $\phi_{ij}\equiv\phi\left(q_{i},x_{j}\right)$,
by solving the (central) finite difference version of Eq. \ref{eq: 2d adjFP}:
\begin{equation}
\begin{split}\left(\frac{2D_{q}}{\Delta q^{2}}+\frac{2D_{x}}{\Delta x^{2}}\right)\phi_{ij}= & D_{q}\frac{\phi_{i+1,j}+\phi_{i-1,j}}{\Delta q^{2}}+D_{x}\frac{\phi_{i,j+1}+\phi_{i,j-1}}{\Delta x^{2}}\\
 & -D_{q}\partial_{q}\beta G_{ij}\frac{\phi_{i+1,j}-\phi_{i-1,j}}{2\Delta q}-D_{x}\partial_{x}\beta G_{ij}\frac{\phi_{i,j+1}-\phi_{i,j-1}}{2\Delta x},
\end{split}
\label{eq: adj FP finite differences}
\end{equation}
where $\partial_{q}\beta G_{ij}$ and $\partial_{x}\beta G_{ij}$
are the gradients of the potential evaluated on the mesh. We set boundary
conditions $\phi_{ij}=1$ for all points $(q<-x_{0},x<-x_{0})$, and
$\phi_{ij}=0$ for all points $(q>x_{0},x>x_{0})$. This definition
of the boundaries assumes that an experienced practitioner will be
able to separate true transitions from mere recrossing events by looking
at the entire trajectory. Additionally, we set reflective boundary
conditions on all remaining points on the border of the mesh,\emph{
i.e.}, $\phi_{i,j}=\phi_{i+1,j}$ for $i=0$ or $i=N-1$, and $\phi_{i,j}=\phi_{i,j+1}$
for $j=0$ or $j=M-1$. We then solve Eq. \ref{eq: adj FP finite differences}
iteratively by initially setting all $\phi_{i,j}$ on the r.h.s. of
the equation inside the boundaries equal to $1/2$. Fig. \ref{fig:3}B
shows an example of the numerical solution $\phi(q,x)$ of Eq. \ref{eq: adj FP finite differences}.

\subsection{Brownian dynamics simulations}

We generated trajectories along $q$ and $x$ using
\begin{equation}
\begin{array}{lcl}
q(t+\Delta t) & = & -\beta\partial_{q}G(q,x)D_{q}\Delta t+(2D_{q}\Delta t)^{1/2}R_{q}(t)\\
x(t+\Delta t) & = & -\beta\partial_{x}G(q,x)D_{x}\Delta t+(2D_{x}\Delta t)^{1/2}R_{x}(t)
\end{array},\label{eq:brownian}
\end{equation}
where $R_{q}(t)$, $R_{x}(t)$ are independent Gaussian random numbers
with zero mean and unit variance, and $\Delta t$ is the time step.
The diffusion coefficient $D_x$ of the molecule is kept constant, and that
of the apparatus $D_{q}$ is varied such that $D_{q}/D_{x}$ ranges
from $10$ to $10^{-2}$. We chose the time step such that $D_{x}\Delta t=5\times10^{-4}$.
Fig. \ref{fig:1}B, shows an example of the measured extension as
a function of time.

\subsection{Estimating the committor from diffusive trajectories}

To calculate the committor directly from a trajectory, we followed
the procedure described by Chodera and Pande.\cite{Chodera2011} For
a trajectory of duration $\tau$, the committor is estimated by 
\begin{equation}
\phi_{\mathrm{traj}}(q)=\frac{\int_{0}^{\tau}\mathrm{d}t\,\delta(q-q(t))c(t)}{\int_{0}^{\tau}\mathrm{d}t\,\delta(q-q(t))},
\end{equation}
where the hitting function $c(t)$ keeps track of whether $q(t)$
hits the folded state before the unfolded one immediately following
time $t$, and assumes a value of unity if so, and zero otherwise. This implies that $c(t)$ uses $q$ only, and does not make use of any indirect information about the hidden variable $x$.  
In practice, we discretized the trajectory $q\left(t\right)$ in space
and time, and considered the resulting discrete chain $j\left(k\right)$,
where $k=0,\ldots,N$ is the time index and $j=0,\ldots,M$ labels
the bins along the extension $q$. The discretized committor estimated
from the trajectory is therefore given by 
\begin{equation}
\phi_{\mathrm{traj}}(i)=\frac{\sum_{k=0}^{N}\delta_{ij\left(k\right)}c\left(k\right)}{\sum_{k=0}^{N}\delta_{ij\left(k\right)}},\label{eq:chodera}
\end{equation}
where $\delta_{ij}$ is the Kronecker delta. Thus, for each bin $i$
(along $q$), $\phi_{\mathrm{traj}}(i)$ is the ratio between the population
committed to the folded state and the total population. In order to
use Eq. \ref{eq:chodera}, we discretized the observed trajectory
in 30 bins between $q=-x_{0}$ and $q=x_{0}$, numerically evaluated
the gradient, and smoothened it with a Savitzky\textendash Golay filter. 

\subsection{Code}

We generated, analyzed, and visualized data with custom code
based on Numpy, \cite{Oliphant} Scipy, \cite{Jones} Ipython,\cite{Perez2007}
Numba\cite{Lam2015} and Matplotlib.\cite{Hunter2007}

\section{Results and discussion}

\begin{figure}
\centering\includegraphics[width=0.8\textwidth]{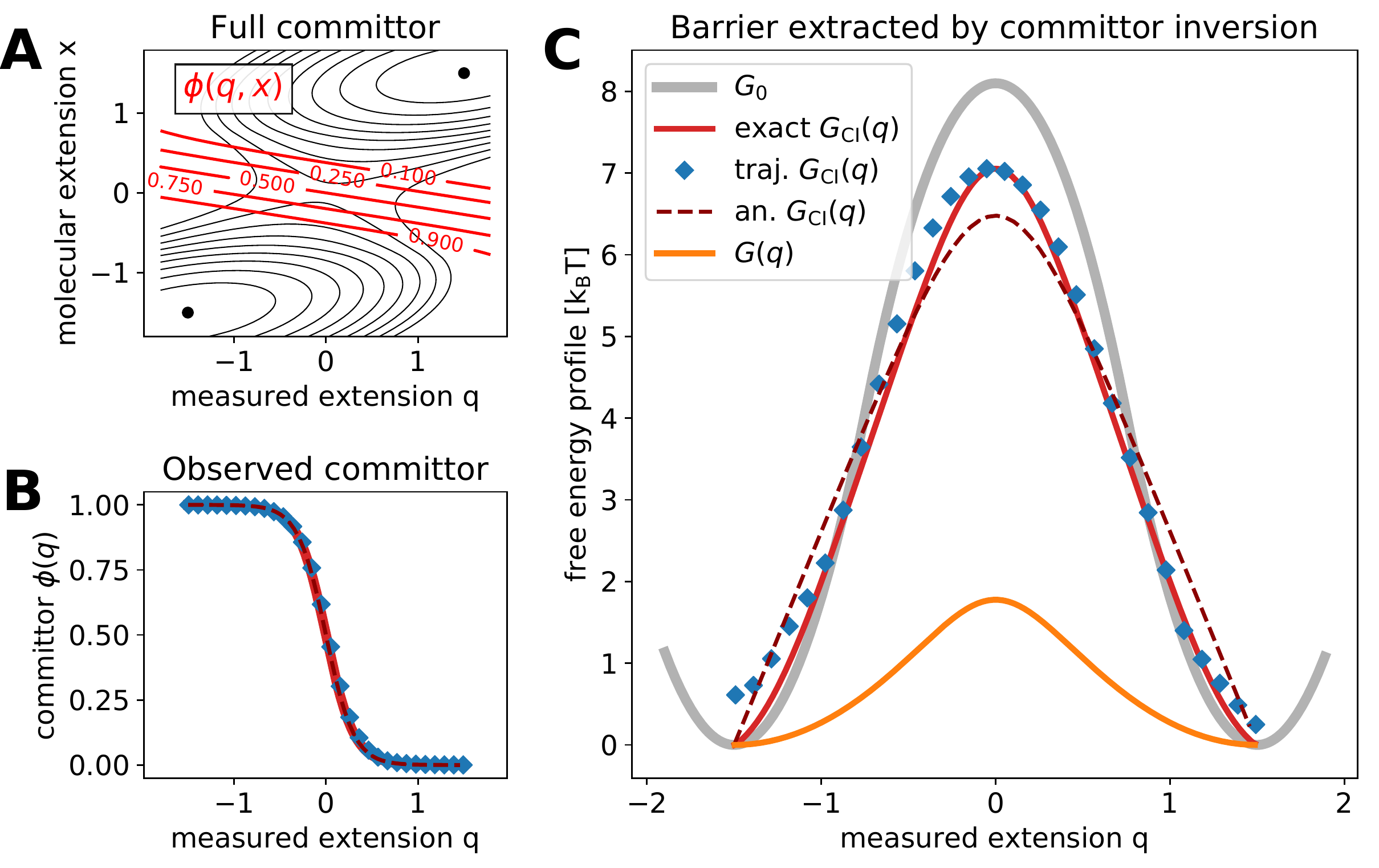}

\caption{\label{fig:3}Molecular free energy profile from committor inversion.
\textbf{ A)} Full two-dimensional committor $\phi\left(q,x\right)$
for the free energy surface $G(q,x)$ and $D_{q}/D_{x}=1$. Isolines of
the free energy surface are shown as black solid lines separated by $1\,k_{\mathrm{B}}T$.
Isolines of the committor are shown as red solid lines. \textbf{B)}
The observed committor $\phi(q)$ is obtained in two independent ways:
from a numerical solution of Onsager's equation (exact, red solid line)
and from Brownian dynamics trajectories (blue diamonds).
The analytic prediction from Eq. \ref{eq:phiq} is shown as a dark red
dashed line. \textbf{C}) The free energy barrier extracted by committor
inversion, $G_{\mathrm{CI}}(q)$, from the exact and trajectory-estimated committor
(red solid line and blue diamonds, respectively). Both barriers are
compared to the hidden molecular profile $G_{o}$ (grey solid line),
to the analytic prediction from Eq. \ref{eq:G_CI_G_q} (dark red dashed),
and to the Boltzmann-inverted free energy profile $G(q)$ (orange solid
line). The free energy surface $G(q,x)$ has parameters similar to those
obtained for the 20TS06/T4 DNA hairpin \cite{Neupane2016}: $\Delta G_{0}^{\ddagger}=8.1\,k_{\mathrm{B}}T$
, $\Delta x^{\ddagger}=1.5\,[x]$, and $\kappa_{l}=2.6\,k_{\mathrm{B}}T/[q]^{2}$,
where $[q]=[x]$ denotes units of length for the extension.}
\end{figure}

We first verified that the observed committor $\phi(q)$ is an equilibrium
conditional average of the full two-dimensional committor $\phi\left(q,x\right)$.
In order to model a typical force spectroscopy experiment, we performed
Brownian dynamics simulations on the two-dimensional potential $G(x,q)$
with $D_{q}/D_{x}=1$ (see Eq. \ref{eq:meancommittor}). For $G_{0}\left(x\right)$,
we used the matched-harmonic potential of Eq. \ref{eq: parabola matched},
and parameters similar to those experimentally obtained for the 20TS06/T4
DNA hairpin.\cite{Neupane2016} We estimated the observed committor
by using Eq. \ref{eq:chodera} from the Brownian dynamics trajectories,
which contained 54 transitions between the minima $q=-x_{o}$ and
$q=x_{o}$. Following a completely independent route, we calculated
$\phi(q,x)$ by numerically solving the Onsager equation (Eqs. \ref{eq: 2d adjFP}
and \ref{eq: adj FP finite differences}, Fig. \ref{fig:3}A), and
obtained $\phi(q)$ as the conditional average in Eq. \ref{eq:meancommittor}.
Fig. \ref{fig:3}B shows these two independent ways to estimate $\phi(q)$,
and compares them to the analytic prediction from Eq. \ref{eq:phiq}
(dashed line). We find that the results from the Brownian dynamics
simulations, accurate numerical calculations, and the analytic prediction are in
excellent agreement.

We used Eq. \ref{eq:G_CI} to invert the mean committor $\phi(q)$
and extracted a free energy profile $G_{\mathrm{CI}}(q)$, both from the accurate numerical
solution and the one estimated from simulated trajectories. Fig. \ref{fig:3}C
reports the results for the 20TS06/T4 DNA hairpin parameters, and
shows a good agreement between the two independent ways for extracting
$G_{\mathrm{CI}}(q)$. 
We compared these results to the potential obtained
from Boltzmann inversion $G(q)$ without deconvolution (orange solid
line). As reported in Ref. \onlinecite{Manuel2015}, the Boltzmann
profile has a much lower barrier, and the two profiles differ significantly.
This  difference indicates that the dynamics along the observed extension $q$
cannot be described as one-dimensional diffusion on $G(q)$.\cite{Chodera2011,Hummer2010}
Interestingly, $G_{\mathrm{CI}}(q)$ is very similar to the hidden molecular
profile $G_{o}$ (grey solid line), and the values of their barriers
differ only by approximately $10\%$, consistent with the results
found in ref. \onlinecite{Manuel2015}. However, Fig. \ref{fig:3}C
also shows that $G_{\mathrm{CI}}\left(q\right)$ is in good agreement with
the analytic approximation from Eq. \ref{eq:G_CI_G_q} (dashed red
line). As can be seen from Eq. \ref{eq:G_CI_G_q}, the analytic expression
does not explicitly depend on $G_{o}(x)$ but only on $G(q)$ and
the stiffnesses of the molecule and linker. This
raises the possibility that the agreement is fortuitous, which motivated
us to further assess the validity of the committor inversion to extract
the hidden molecular profile for a large number of scenarios.

\begin{figure}
\includegraphics[width=0.4\textwidth]{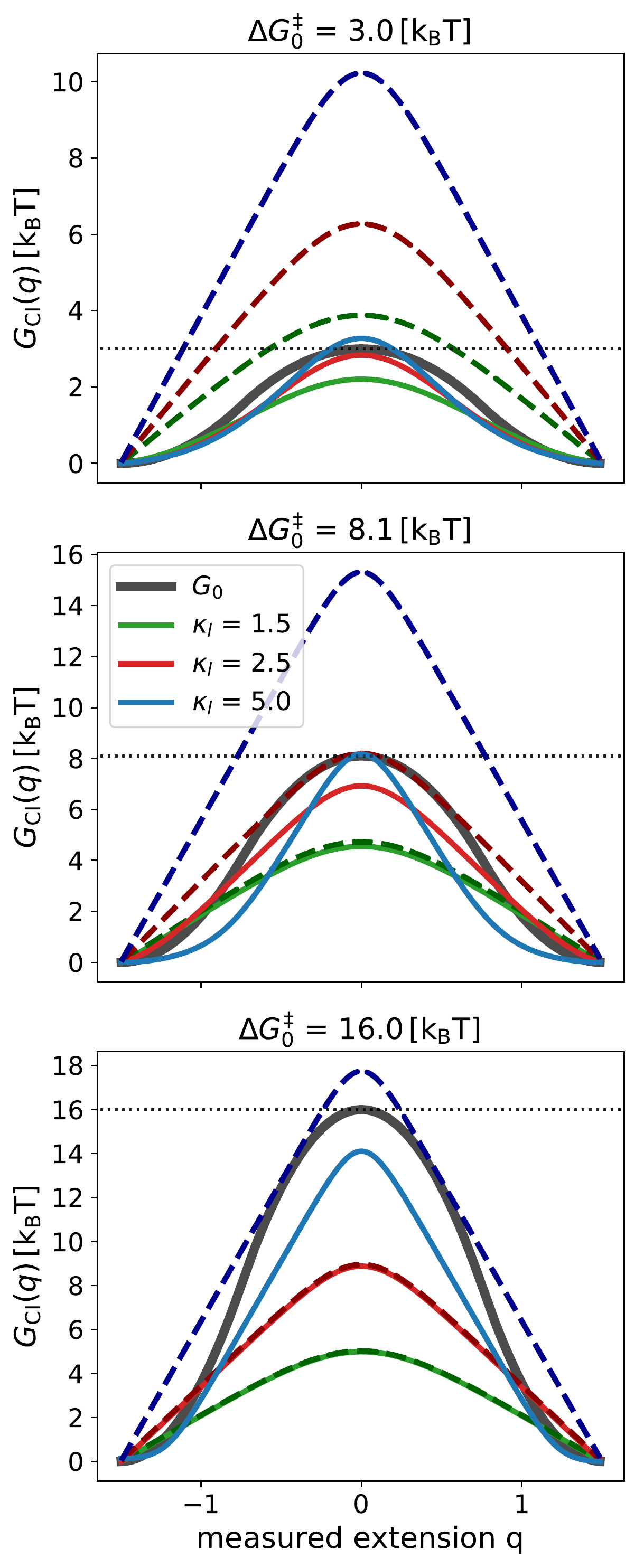}

\caption{\label{fig:4}Free energy barrier extracted by committor inversion,
$G_{\mathrm{CI}}(q)$, using the observed
committor (solid lines) and
the analytic approximation of Eq. \ref{eq:G_CI_G_q} (dashed lines and darker shade of color).
We varied the height of the hidden molecular barrier ($\Delta G_{o}^{\ddagger}=3$,
$8.1$, and $16\,k_{\mathrm{B}}T$, from top to bottom) and the
linker stiffness ($\kappa_{l}=1.5$, $2.5$, and $5\,k_{\mathrm{B}}T/[q]^{2}$,
green, red, and blue lines, respectively). We used the matched-harmonic
free energy function from Eq. \ref{eq: parabola matched} with $\Delta x^{\ddagger}=1.5\,[x]$.
For reference, each panel shows the respective hidden molecular profile $G_{o}$ (grey solid line).}
\end{figure}

We investigated how well the free energy profile obtained by inversion
of the observed committor, $G_{\mathrm{CI}}(q)$, reproduces the hidden molecular
profile, $G_{o}$, for a number of cases of practical interest. Having
shown that the observed committor is accurately reproduced by numerical
solutions of Onsager's equation, we used the latter to systematically
investigate the influence of parameters of our model. In Fig. \ref{fig:4}
we report the dependence of the exact $G_{\mathrm{CI}}(q)$ on the linker stiffness
(solid lines) and on the height of the hidden molecular barrier. The
results show that the accuracy of predicting $G_{0}$ depends on all
the examined parameters. For instance, using the linker stiffness
$\kappa_{l}=1.5\,k_{\mathrm{B}}T/[q]^{2}$, where $[x]=[q]$ indicates
the units of extension, guarantees acceptable results for $\Delta G_{0}^{\ddagger}=3\,k_{\mathrm{B}}T$
but works rather poorly for larger barriers, as can be seen for $\Delta G_{0}^{\ddagger}=8.1\,k_{\mathrm{B}}T$.
For $\Delta G_{0}^{\ddagger}=16\,k_{\mathrm{B}}T$, only a very stiff
linker gives an acceptable reconstruction.

We tested the validity of the analytic approximation Eq. \ref{eq:G_CI_G_q}
(dashed lines in Fig. \ref{fig:4}). We found that Eq. \ref{eq:G_CI_G_q}
reproduces accurately the exact solutions for sufficiently large molecular
barriers ($\geq5\,k_{\mathrm{B}}T$) and for soft linkers. Since the
analytic formula does not contain explicit information about the hidden
molecular profile (only information about the potential wells),
whenever this approximation accurately reproduces $G_{\mathrm{CI}}(q)$ the
reconstruction of $G_{o}$ by committor inversion is likely to be
invalid. Notably, in these cases, the barrier height obtained by committor
inversion is systematically lower than the molecular barrier. 

\begin{figure}
\centering\includegraphics[width=1\textwidth]{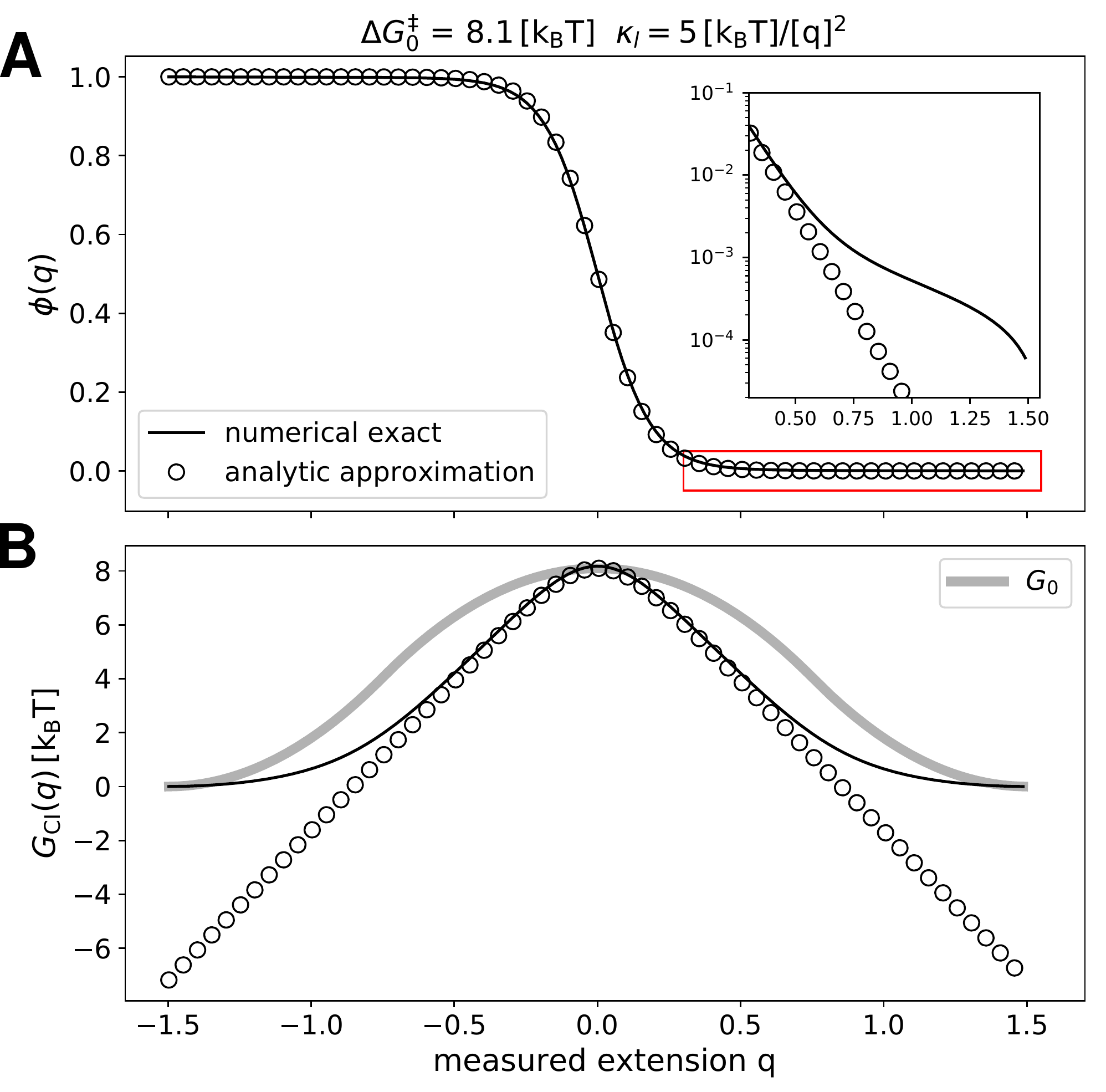}

\caption{\label{fig:5}\textbf{A}) Observed committor $\phi(q)$ and \textbf{B})
barrier $G_{\mathrm{CI}}(q)$ obtained by committor inversion for the exact
numerical solution (solid line) and the analytic approximation of
Eq. \ref{eq:G_CI_G_q} (empty circles \textendash{} to highlight agreement
with exact solution). We used parameters $\Delta G_{o}^{\ddagger}=8.1\,k_{\mathrm{B}}T$
and $\kappa_{l}=5\,k_{\mathrm{B}}T/[q]^{2}$. The inset zooms in on
$\phi(q)$ in the range highlighted by red square box. For reference
$G_{o}$ is shown in the bottom panel as a gray solid line. All curves
in B) are aligned on the barrier top. A seemingly insignificant difference
on the committors shown in A leads to an overestimation of the barrier
height by a factor 2.}
\end{figure}

We then investigated the cases in which the analytic approximation
does not correctly reproduce the free energy profile obtained by the
exact numerical solution of the committor inversion, even when the
barrier seems sufficiently high. This can be seen, for instance, in
Fig. \ref{fig:4} for $\Delta G_{o}^{\ddagger}=8.1\,k_{\mathrm{B}}T$
and $\kappa_{l}=5\,{k_{\mathrm{B}}T}/[q]^{2}$. We compared the exact
$\phi(q)$ to the analytic prediction from Eq. \ref{eq:phiq} (Fig.
\ref{fig:5}). The two quantities are in striking agreement over most
of the reaction coordinate range, and deviate only close to the minima
by exponentially small amounts, which cause the slopes $\phi'(q)$
of the two curves to be different (Fig. \ref{fig:5} inset in panel
A). These differences in $\phi'(q)$ are amplified by the logarithm
in Eq. \ref{eq:G_CI}, causing large errors in the inverted free energy
barrier $G_{\mathrm{CI}}(q)$ at the well bottom (Fig. \ref{fig:5} panel B)
that lead to systematic underestimation of the barrier height. In
fact, Eq. \ref{eq:phiq} accurately reproduces the exact $G_{\mathrm{CI}}(q)$
around the top of the barrier, but poorly describes how $G_{\mathrm{CI}}(q)$
approaches its minima. Fig. \ref{fig:5} indicates that the mean committor
close to the stable states encodes crucial information about the height
of the extracted barrier, and must be estimated with very high precision. 
This requirement poses a major challenge for practical
attempts to reconstruct barriers by committor inversion.

\begin{figure}
\includegraphics[width=12cm]{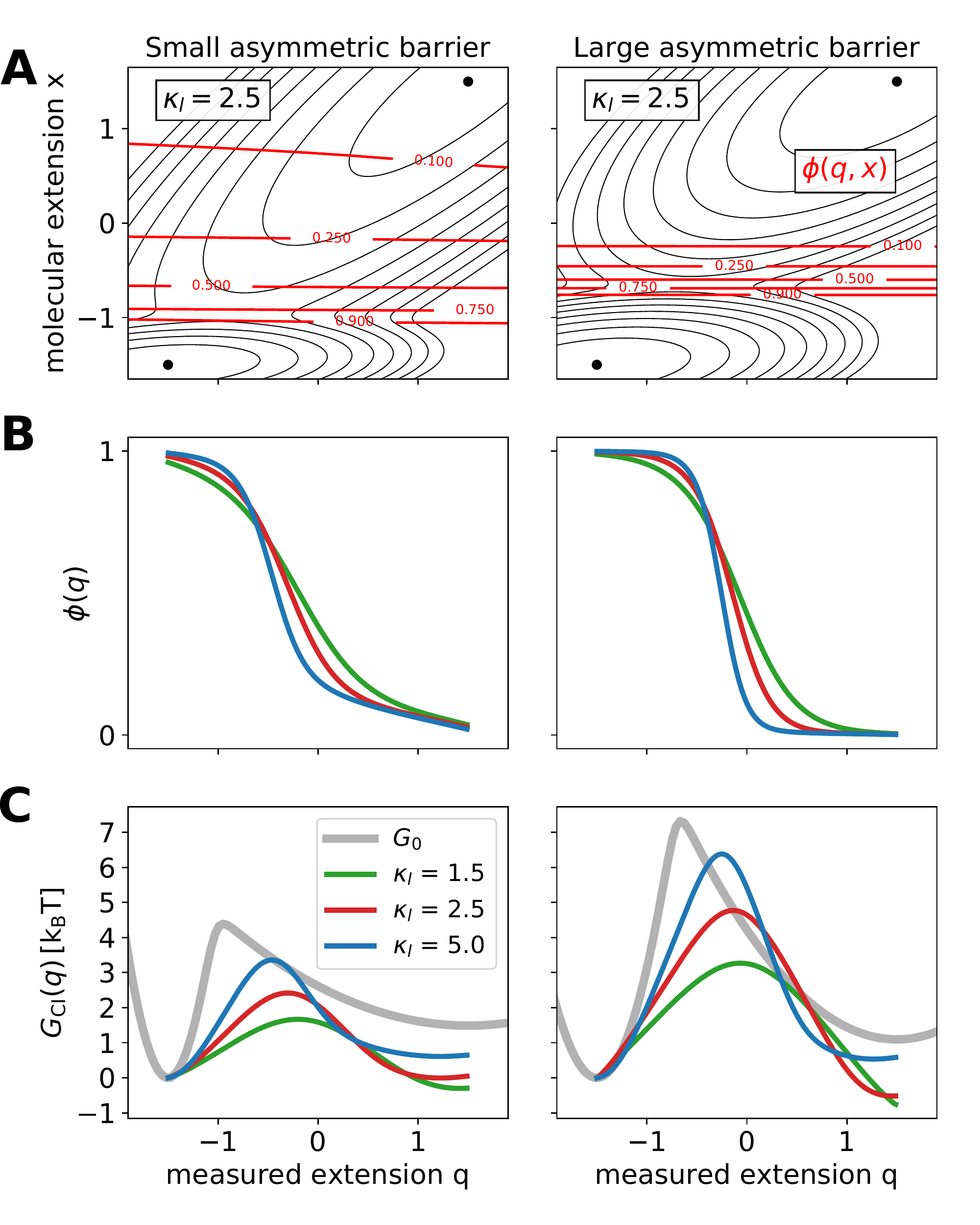}\caption{\label{fig:6}
  Asymmetric molecular barriers and free energy profiles extracted
by committor inversion.\textbf{ A)} Full two-dimensional committor
$\phi\left(q,x\right)$ for the free energy surfaces $G(q,x)$ with an
asymmetric molecular free energy $G_{o}^{\mathrm{asym}}$ (Eq. \ref{eq:Gasym}),
$\kappa_{l}=2.5\,{k_{\mathrm{B}}T}/[q]^{2}$, and $D_{q}/D_{x}=1$.
Isolines of the free energy surfaces are shown as black solid lines separated
by $1\,k_{\mathrm{B}}T$. Isolines of the committor are shown
as red solid lines. We considered free energy surfaces with a small and
a large barrier, left and right, respectively. \textbf{{B)}}{{}
The observed committor $\phi(q)$ is obtained by numerically solving
Onsager's equation. $\phi(q)$ is shown for different linker stiffness
($1.5$, $2.5$, and }\textbf{{$5.0$ }}{${k_{\mathrm{B}}T}/[q]^{2}$).}\textbf{
C)} Free energy profiles 
$G_{\mathrm{\mathrm{CI}}}(q)$ extracted by committor inversion. 
The respective hidden molecular profiles
$G_{o}^{\mathrm{asym}}$ are shown as grey solid lines.}
\end{figure}

We also investigated how well committor inversion allows one to estimate
the shape of the hidden molecular barrier for asymmetric molecular
energy profiles. In Fig. \ref{fig:6}A, we show the two-dimensional
free energy surface $G(q,x)$ and two-dimensional committor for small (left)
and large (right) asymmetric barriers. On both free energy surfaces, the
well to the left of the barrier is narrower and deeper than that to
the right of the barrier. The asymmetry of the barrier is reflected
in the full two-dimensional committor. In fact, the committor isoline
of 0.5 is not located at the barrier-top but displaced 
towards the
shallower state, whereas points on the top of the barrier are actually
highly committed. $\phi(q)$ obtained by solving the Onsager's equation
and the profiles extracted by committor inversion are shown in Fig.
\ref{fig:6} (B and C, respectively) for different linker stiffness.
The comparison to the hidden molecular free energy (gray line) shows that
the asymmetry of the molecular free energy can be captured only qualitatively
under the conditions used here. Indeed, the accuracy in determining
both the location of the barrier top and the relative stability
of the two states depends on the stiffness of the linker, and improves
with increasingly stiffer linkers. For low linker stiffness, the barrier
from committor inversion is systematically lower and closer to $q=0$ than in $G_{o}^{\mathrm{asym}}$.

\begin{figure}
\includegraphics[width=1\textwidth]{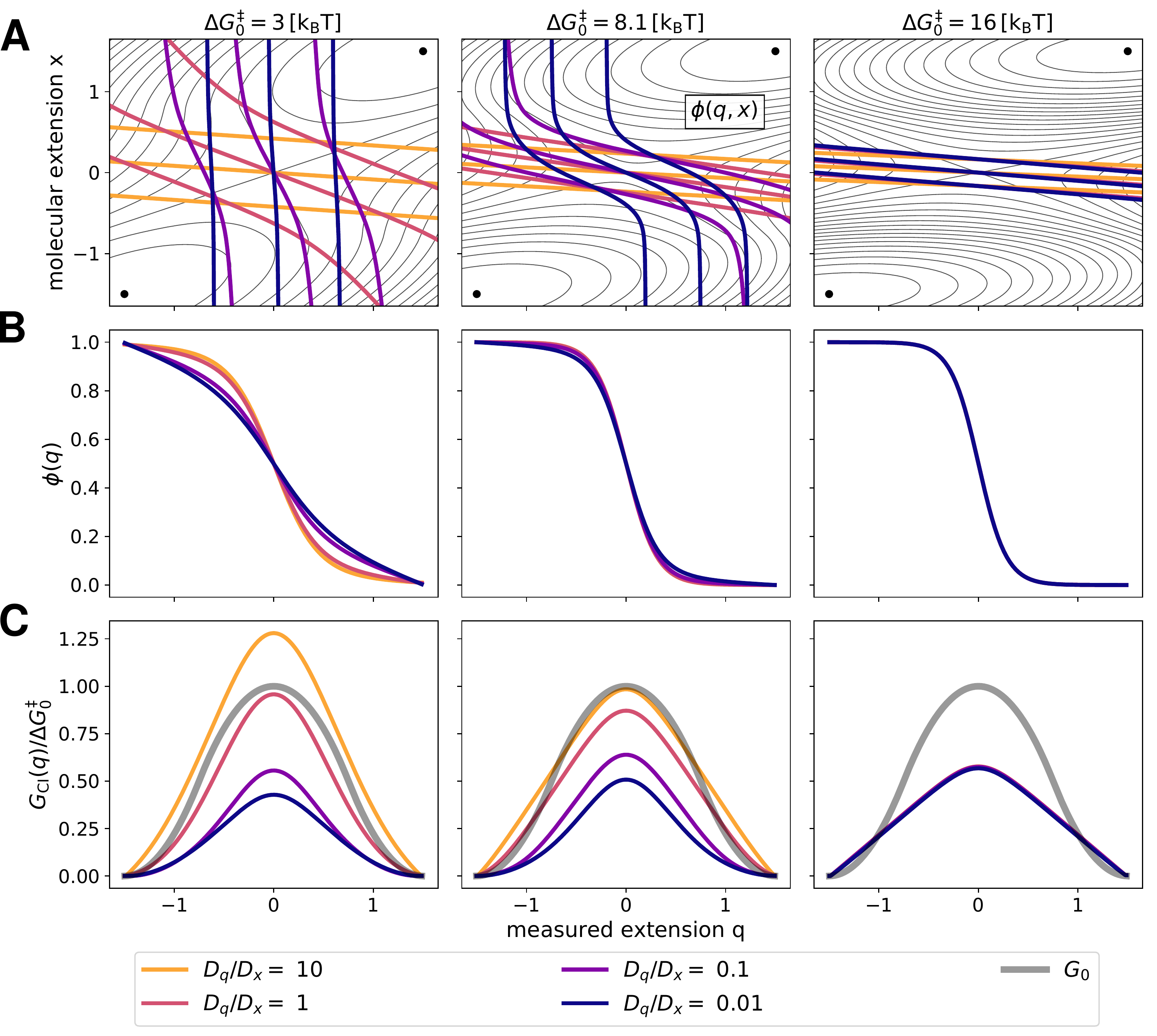}

\caption{\label{fig:7} Effect of diffusion anisotropy on committors
  and free energy
barriers extracted by committor inversion. \textbf{A}) Free energy surfaces
and corresponding full committor $\phi\left(q,x\right)$ calculated
for different diffusion anisotropies $D_{q}/D_{x}$ and increasing
values of the molecular barrier height 
($\Delta G_{o}^{\ddagger}=3$,
$8.1$, and $16\,k_{\mathrm{B}}T$, from left to right) with $\kappa_{l}=2.6\,k_{\mathrm{B}}T/[q]^{2}$
in each case. Isolines of the free energy surfaces are shown as black solid
lines separated by $1\,k_{\mathrm{B}}T$. Isolines of the
committor corresponding to $0.2$, $0.5$, and $0.8$ are shown as
colored solid lines. \textbf{B}) Corresponding observed committor
$\phi\left(q\right)$ as a function of diffusion anisotropy $D_{q}/D_{x}$.
\textbf{C}) Barrier $G_{\mathrm{CI}}(q)$ obtained by inversion of the observed
committors shown in B. The hidden molecular barrier $G_{0}$ is
reported as a solid grey line. In each panel of C, free energies are measured
in units of the corresponding value of $\Delta G_{0}^{\ddagger}$.
Color code for diffusion anisotropy $D_{q}/D_{x}$: $10$ orange,
$1$ red, $0.1$ purple, $0.01$ blue. In the rightmost panel, the
curves for $G_{\mathrm{CI}}(q)$ are superimposed. }
\end{figure}

Finally, we studied the effects of diffusion anisotropy on $\phi\left(q,x\right)$,
$\phi(q)$, and $G_{\mathrm{CI}}(q)$ by numerically obtaining the solutions
of Eq. \ref{eq: 2d adjFP} as a function of the ratio $D_{q}/D_{x}$
over four orders of magnitude, using different molecular barrier heights
$\Delta G_{0}^{\ddagger}$ (Fig. \ref{fig:7}). For small molecular
barriers, reducing $D_{q}$ (corresponding to slower diffusion of
the force probe attached to the molecule) induces a ``rotation''
of the full committor $\phi\left(q,x\right)$ around the barrier (Fig.
\ref{fig:7} A). For $D_{q}/D_{x}=10$, the isolines of the committor
are almost parallel to the $q$-axis, indicating transitions that
are dominated by the dynamics along $x$ (Fig. \ref{fig:7} A orange
lines). As $D_{q}$ decreases the isolines rotate, until they are
perpendicular to the $q$-axis for very small $D_{q}$, indicating
that transitions are dominated by the much slower dynamics along $q$
(blue lines). This phenomenon is most clearly observed for $\Delta G_{0}^{\ddagger}=3\,{k_{\mathrm{B}}T}$.
For $\Delta G_{0}^{\ddagger}=8.1\,{k_{\mathrm{B}}T}$ larger values
of diffusion anisotropy are required to induce rotations of the isolines
of the full committor, which are mostly suppressed on the barrier.
For the largest barrier, $\Delta G_{0}^{\ddagger}=16\,k_{\mathrm{B}}T$,
diffusion anisotropy has no sizable effect on the committor, which
is completely determined by the free energy surface.

Consequently, diffusion anisotropy has a significant impact on $\phi(q)$
and $G_{\mathrm{CI}}(q)$ for low and medium barrier heights, but no effect
for very large ones. For decreasing values of $D_{q}/D_{x}$, the
barrier reconstructed by committor inversion tends to increasingly
underestimate the hidden molecular barrier $\Delta G_{0}^{\ddagger}$.
This observation represents a further challenge for practical applications
of the committor inversion method to experiments, since the probe
diffusion may well be much slower than the molecular diffusion (depending
on the molecule being studied and the design of the probe). As shown
already in Fig. \ref{fig:5}, small variations in the observed committor
arising from diffusion anisotropy can have dramatic effects on the
barrier height estimated by inversion (Fig. \ref{fig:7} B-C, middle
panel).

\section{Concluding remarks}

We have assessed the validity of committor inversion
to extract molecular free energy profiles from single-molecule force
spectroscopy experiments. Within the framework of a two-dimensional
model for the coupled dynamics of the molecular and measured extensions,
we obtained approximate analytic expressions for the measured committor
and the extracted free energy profile from its inversion. We compared
these analytic results with those obtained from Brownian dynamics
simulations and accurate numerical solutions of the Onsager equation
for various linker stiffness values and molecular barrier heights.
We found that for isotropic diffusion the committor inversion gives
reasonable results for small and medium-high barriers, and that the
accuracy depends on the stiffness of the linker. When the apparatus
diffuses much more slowly than the molecule, or when the barrier is
high, the reconstruction is far less accurate. We have also shown
that due to the logarithms in the inversion formula, even exponentially
small inaccuracies in the observed committor lead to large errors
in the barrier height of the reconstructed molecular free energy
profile. This may represent a serious challenge for practical application
of the committor inversion approach. Although in some situations molecular free
energy profiles estimated by committor inversion from single-molecule
experiments can be informative, systematically ascertaining their validity
is challenging and they should in general be regarded with caution. 

\section{Acknowledgments}

R.C., G.H., and P.C. acknowledge the support of the Max Planck Society.
P.C. was also supported by Colciencias, University of Antioquia, and
Ruta N, Colombia. A.S. was supported by the Intramural Research Program
of the National Institute of Diabetes and Digestive and Kidney Diseases
of the National Institutes of Health. M. T. W. acknowledges support
from the John Simon Guggenheim Foundation.

\bibliographystyle{apsrev}
\bibliography{biblio}

\begin{thebibliography}{18}
\expandafter\ifx\csname natexlab\endcsname\relax\def\natexlab#1{#1}\fi
\expandafter\ifx\csname bibnamefont\endcsname\relax
  \def\bibnamefont#1{#1}\fi
\expandafter\ifx\csname bibfnamefont\endcsname\relax
  \def\bibfnamefont#1{#1}\fi
\expandafter\ifx\csname citenamefont\endcsname\relax
  \def\citenamefont#1{#1}\fi
\expandafter\ifx\csname url\endcsname\relax
  \def\url#1{\texttt{#1}}\fi
\expandafter\ifx\csname urlprefix\endcsname\relax\def\urlprefix{URL }\fi
\providecommand{\bibinfo}[2]{#2}
\providecommand{\eprint}[2][]{\url{#2}}

\bibitem[{\citenamefont{Greenleaf et~al.}(2007)\citenamefont{Greenleaf,
  Woodside, and Block}}]{Greenleaf2007}
\bibinfo{author}{\bibfnamefont{W.~J.} \bibnamefont{Greenleaf}},
  \bibinfo{author}{\bibfnamefont{M.~T.} \bibnamefont{Woodside}},
  \bibnamefont{and} \bibinfo{author}{\bibfnamefont{S.~M.} \bibnamefont{Block}},
  \bibinfo{journal}{{Annu. Rev. Biophys. Biomol. Struct.}}
  \textbf{\bibinfo{volume}{36}}, \bibinfo{pages}{171} (\bibinfo{year}{2007}).

\bibitem[{\citenamefont{Neuman and Nagy}(2008)}]{Neuman2008}
\bibinfo{author}{\bibfnamefont{K.~C.} \bibnamefont{Neuman}} \bibnamefont{and}
  \bibinfo{author}{\bibfnamefont{A.}~\bibnamefont{Nagy}},
  \bibinfo{journal}{{Nat. Methods}} \textbf{\bibinfo{volume}{5}},
  \bibinfo{pages}{491} (\bibinfo{year}{2008}).

\bibitem[{\citenamefont{Onsager}(1938)}]{Onsager1938}
\bibinfo{author}{\bibfnamefont{L.}~\bibnamefont{Onsager}},
  \bibinfo{journal}{Phys. Rev.} \textbf{\bibinfo{volume}{54}},
  \bibinfo{pages}{554} (\bibinfo{year}{1938}), ISSN \bibinfo{issn}{0031-899X},
  \urlprefix\url{https://link.aps.org/doi/10.1103/PhysRev.54.554}.

\bibitem[{\citenamefont{Chodera and Pande}(2011)}]{Chodera2011}
\bibinfo{author}{\bibfnamefont{J.~D.} \bibnamefont{Chodera}} \bibnamefont{and}
  \bibinfo{author}{\bibfnamefont{V.~S.} \bibnamefont{Pande}},
  \bibinfo{journal}{Phys. Rev. Lett.} \textbf{\bibinfo{volume}{107}},
  \bibinfo{pages}{098102} (\bibinfo{year}{2011}), ISSN
  \bibinfo{issn}{0031-9007}, \eprint{1105.0710},
  \urlprefix\url{https://link.aps.org/doi/10.1103/PhysRevLett.107.098102}.

\bibitem[{\citenamefont{Manuel et~al.}(2015)\citenamefont{Manuel, Lambert, and
  Woodside}}]{Manuel2015}
\bibinfo{author}{\bibfnamefont{A.~P.} \bibnamefont{Manuel}},
  \bibinfo{author}{\bibfnamefont{J.}~\bibnamefont{Lambert}}, \bibnamefont{and}
  \bibinfo{author}{\bibfnamefont{M.~T.} \bibnamefont{Woodside}},
  \bibinfo{journal}{Proc. Natl. Acad. Sci.} \textbf{\bibinfo{volume}{112}},
  \bibinfo{pages}{7183} (\bibinfo{year}{2015}), ISSN \bibinfo{issn}{0027-8424},
  \urlprefix\url{http://www.pnas.org/lookup/doi/10.1073/pnas.1419490112}.

\bibitem[{\citenamefont{Woodside et~al.}(2006)\citenamefont{Woodside, Anthony,
  Behnke-Parks, Larizadeh, Herschlag, and Block}}]{Woodside2006}
\bibinfo{author}{\bibfnamefont{M.~T.} \bibnamefont{Woodside}},
  \bibinfo{author}{\bibfnamefont{P.~C.} \bibnamefont{Anthony}},
  \bibinfo{author}{\bibfnamefont{W.~M.} \bibnamefont{Behnke-Parks}},
  \bibinfo{author}{\bibfnamefont{K.}~\bibnamefont{Larizadeh}},
  \bibinfo{author}{\bibfnamefont{D.}~\bibnamefont{Herschlag}},
  \bibnamefont{and} \bibinfo{author}{\bibfnamefont{S.~M.} \bibnamefont{Block}},
  \bibinfo{journal}{Science (80-. ).} \textbf{\bibinfo{volume}{314}},
  \bibinfo{pages}{1001} (\bibinfo{year}{2006}), ISSN \bibinfo{issn}{0036-8075},
  \urlprefix\url{http://www.sciencemag.org/cgi/doi/10.1126/science.1133601}.

\bibitem[{\citenamefont{Hummer and Szabo}(2010)}]{Hummer2010}
\bibinfo{author}{\bibfnamefont{G.}~\bibnamefont{Hummer}} \bibnamefont{and}
  \bibinfo{author}{\bibfnamefont{A.}~\bibnamefont{Szabo}},
  \bibinfo{journal}{Proc. Natl. Acad. Sci.} \textbf{\bibinfo{volume}{107}},
  \bibinfo{pages}{21441} (\bibinfo{year}{2010}), ISSN
  \bibinfo{issn}{0027-8424}.

\bibitem[{\citenamefont{Yu et~al.}(2017)\citenamefont{Yu, Siewny, Edwards,
  Sanders, and Perkins}}]{Yu2017}
\bibinfo{author}{\bibfnamefont{H.}~\bibnamefont{Yu}},
  \bibinfo{author}{\bibfnamefont{M.~G.~W.} \bibnamefont{Siewny}},
  \bibinfo{author}{\bibfnamefont{D.~T.} \bibnamefont{Edwards}},
  \bibinfo{author}{\bibfnamefont{A.~W.} \bibnamefont{Sanders}},
  \bibnamefont{and} \bibinfo{author}{\bibfnamefont{T.~T.}
  \bibnamefont{Perkins}}, \bibinfo{journal}{Science (80-. ).}
  \textbf{\bibinfo{volume}{355}}, \bibinfo{pages}{945} (\bibinfo{year}{2017}),
  ISSN \bibinfo{issn}{0036-8075},
  \urlprefix\url{http://www.sciencemag.org/lookup/doi/10.1126/science.aah7124}.

\bibitem[{\citenamefont{Makarov}(2014)}]{Makarov2014}
\bibinfo{author}{\bibfnamefont{D.~E.} \bibnamefont{Makarov}},
  \bibinfo{journal}{The Journal of Chemical Physics}
  \textbf{\bibinfo{volume}{141}}, \bibinfo{pages}{241103}
  (\bibinfo{year}{2014}), ISSN \bibinfo{issn}{0021-9606},
  \urlprefix\url{http://aip.scitation.org/doi/10.1063/1.4904895}.

\bibitem[{\citenamefont{Hyeon et~al.}(2008)\citenamefont{Hyeon, Morrison, and
  Thirumalai}}]{Hyeon2008}
\bibinfo{author}{\bibfnamefont{C.}~\bibnamefont{Hyeon}},
  \bibinfo{author}{\bibfnamefont{G.}~\bibnamefont{Morrison}}, \bibnamefont{and}
  \bibinfo{author}{\bibfnamefont{D.}~\bibnamefont{Thirumalai}},
  \bibinfo{journal}{Proceedings of the National Academy of Sciences}
  \textbf{\bibinfo{volume}{105}}, \bibinfo{pages}{9604} (\bibinfo{year}{2008}),
  ISSN \bibinfo{issn}{0027-8424},
  \urlprefix\url{http://www.pnas.org/cgi/doi/10.1073/pnas.0802484105}.

\bibitem[{\citenamefont{Cossio et~al.}(2015)\citenamefont{Cossio, Hummer, and
  Szabo}}]{Cossio2015}
\bibinfo{author}{\bibfnamefont{P.}~\bibnamefont{Cossio}},
  \bibinfo{author}{\bibfnamefont{G.}~\bibnamefont{Hummer}}, \bibnamefont{and}
  \bibinfo{author}{\bibfnamefont{A.}~\bibnamefont{Szabo}},
  \bibinfo{journal}{Proc. Natl. Acad. Sci.} \textbf{\bibinfo{volume}{112}},
  \bibinfo{pages}{14248} (\bibinfo{year}{2015}), ISSN
  \bibinfo{issn}{0027-8424},
  \urlprefix\url{http://www.pnas.org/lookup/doi/10.1073/pnas.1519633112}.

\bibitem[{\citenamefont{Cossio et~al.}(2018)\citenamefont{Cossio, Hummer, and
  Szabo}}]{Cossio2018}
\bibinfo{author}{\bibfnamefont{P.}~\bibnamefont{Cossio}},
  \bibinfo{author}{\bibfnamefont{G.}~\bibnamefont{Hummer}}, \bibnamefont{and}
  \bibinfo{author}{\bibfnamefont{A.}~\bibnamefont{Szabo}}, \bibinfo{journal}{J.
  Chem. Phys.} \textbf{\bibinfo{volume}{148}}, \bibinfo{pages}{123309}
  (\bibinfo{year}{2018}), ISSN \bibinfo{issn}{0021-9606},
  \urlprefix\url{http://dx.doi.org/10.1063/1.5004767
  http://aip.scitation.org/doi/10.1063/1.5004767}.

\bibitem[{\citenamefont{Oliphant}(2015)}]{Oliphant}
\bibinfo{author}{\bibfnamefont{T.~E.} \bibnamefont{Oliphant}},
  \emph{\bibinfo{title}{{Guide to NumPy}}} (\bibinfo{publisher}{CreateSpace
  Independent Publishing Platform}, \bibinfo{address}{USA},
  \bibinfo{year}{2015}), \bibinfo{edition}{2nd} ed., ISBN
  \bibinfo{isbn}{151730007X, 9781517300074}.

\bibitem[{\citenamefont{Jones et~al.}()\citenamefont{Jones, Oliphant, Peterson,
  and Others}}]{Jones}
\bibinfo{author}{\bibfnamefont{E.}~\bibnamefont{Jones}},
  \bibinfo{author}{\bibfnamefont{T.}~\bibnamefont{Oliphant}},
  \bibinfo{author}{\bibfnamefont{P.}~\bibnamefont{Peterson}}, \bibnamefont{and}
  \bibinfo{author}{\bibnamefont{Others}}, \emph{\bibinfo{title}{{SciPy: Open
  source scientific tools for Python}}}, \urlprefix\url{http://www.scipy.org/}.

\bibitem[{\citenamefont{Perez and Granger}(2007)}]{Perez2007}
\bibinfo{author}{\bibfnamefont{F.}~\bibnamefont{Perez}} \bibnamefont{and}
  \bibinfo{author}{\bibfnamefont{B.~E.} \bibnamefont{Granger}},
  \bibinfo{journal}{Comput. Sci. Eng.} \textbf{\bibinfo{volume}{9}},
  \bibinfo{pages}{21} (\bibinfo{year}{2007}), ISSN \bibinfo{issn}{1521-9615},
  \urlprefix\url{http://ieeexplore.ieee.org/lpdocs/epic03/wrapper.htm?arnumber=4160251}.

\bibitem[{\citenamefont{Lam et~al.}(2015)\citenamefont{Lam, Pitrou, and
  Seibert}}]{Lam2015}
\bibinfo{author}{\bibfnamefont{S.~K.} \bibnamefont{Lam}},
  \bibinfo{author}{\bibfnamefont{A.}~\bibnamefont{Pitrou}}, \bibnamefont{and}
  \bibinfo{author}{\bibfnamefont{S.}~\bibnamefont{Seibert}}, in
  \emph{\bibinfo{booktitle}{Proc. Second Work. LLVM Compil. Infrastruct. HPC -
  LLVM '15}} (\bibinfo{publisher}{ACM Press}, \bibinfo{address}{New York, New
  York, USA}, \bibinfo{year}{2015}), pp. \bibinfo{pages}{1--6}, ISBN
  \bibinfo{isbn}{9781450340052},
  \urlprefix\url{http://dl.acm.org/citation.cfm?doid=2833157.2833162}.

\bibitem[{\citenamefont{Hunter}(2007)}]{Hunter2007}
\bibinfo{author}{\bibfnamefont{J.~D.} \bibnamefont{Hunter}},
  \bibinfo{journal}{Comput. Sci. Eng.} \textbf{\bibinfo{volume}{9}},
  \bibinfo{pages}{90} (\bibinfo{year}{2007}), ISSN \bibinfo{issn}{1521-9615},
  \urlprefix\url{http://ieeexplore.ieee.org/lpdocs/epic03/wrapper.htm?arnumber=4160265}.

\bibitem[{\citenamefont{Neupane and Woodside}(2016)}]{Neupane2016}
\bibinfo{author}{\bibfnamefont{K.}~\bibnamefont{Neupane}} \bibnamefont{and}
  \bibinfo{author}{\bibfnamefont{M.~T.} \bibnamefont{Woodside}},
  \bibinfo{journal}{Biophys. J.} \textbf{\bibinfo{volume}{111}},
  \bibinfo{pages}{283} (\bibinfo{year}{2016}), ISSN \bibinfo{issn}{15420086},
  \urlprefix\url{http://dx.doi.org/10.1016/j.bpj.2016.06.011}.

\end{thebibliography}

\end{document}